 \documentclass{jfm}

\lefttitle{B. Dhote et al.}
\righttitle{Journal of Fluid Mechanics}

\title{Capillary effects on preferential orientation of floaters in gravity waves}

\author{Basile Dhote\aff{1}, Ewen Le~Ster\aff{1}, Wietze Herreman\aff{1}, Fr\'ed\'eric Moisy\aff{1}}
\affiliation{\aff{1}Universit\'e Paris-Saclay, CNRS, Laboratoire FAST, 91405 Orsay, France}

\usepackage{graphicx}
\usepackage{dcolumn}
\usepackage{bm}
\usepackage{ulem}
\usepackage{comment}
\usepackage{esint}
\usepackage[toc,page]{appendix}
\usepackage{amsmath}

\newcommand{\psib}{\overline{\psi}}
\newcommand{\hb}{\overline{h}}

\newcommand{\bs}{\boldsymbol}

\newcommand{\br}{\mathbf{r}}
\newcommand{\brc}{\mathbf{r}_c}

\newcommand{\wt}{\widetilde}

\newcommand{\xt}{\widetilde{x}}
\newcommand{\yt}{\widetilde{y}}
\newcommand{\zt}{\widetilde{z}}
\newcommand{\xbc}{\overline{x}_c}

\newcommand{\spsi}{s_{\psi}}
\newcommand{\sbpsi}{\overline{s}_{\psi}}
\newcommand{\cpsi}{c_{\psi}}
\newcommand{\cbpsi}{\overline{c}_{\psi}}

\newcommand{\Ssub}{S_{\text{wet}}}

\newcommand{\Vdisp}{V_{\text{disp}}}

\newcommand{\lc}{\ell_c}

\newcommand{\lx}{l_x}
\newcommand{\ly}{l_y}
\newcommand{\lz}{l_z}

\newcommand{\zetapb}{\overline{\zeta}_p}

\newcommand{\bdS}{\boldsymbol{dS}}

\begin{document}

\maketitle

\begin{abstract}
We study the influence of capillary effects on the motion of thin elastic plates denser than water drifting in propagating surface gravity waves. Such floaters experience a mean angular drift that rotates them toward two preferential orientations: parallel to the direction of wave propagation (longitudinal) or parallel to the wave crests (transverse). We develop a diffractionless model (Froude-Krylov approximation) to compute the mean yaw moment acting on floaters with arbitrary bending rigidity, small relative to the wavelength. Capillary forces are incorporated through a quasi-static volume formulation based on the fluid volume displaced by the floater and its meniscus. The model predicts that the preferential orientation is governed by the non-dimensional parameter $F = kL_x^2/\hb$ recently introduced in Herreman {\it et al.} ({\it J. Fluid Mech.}, vol.~999, 2024, art.~A92), where $k$ is the wavenumber, $L_x$ the floater length, and $\hb$ the equilibrium immersion depth, provided that $\hb$ accounts for capillary effects. The orientation depends on how $F$ compares to a critical value $F_c$, which is a function of the ratio of the flexural length to the floater length. These predictions are in good agreement with experiments performed with thin metal rectangular plates of various length, width and thickness.
\end{abstract}

\section{Introduction}

Capillarity plays a central role in the dynamics of small-scale objects floating at liquid interfaces, and its influence has been extensively investigated in the context of interfacial bio-locomotion. Some insects exploit surface deformations to navigate on water, such as water striders, which remain supported without penetrating the interface \citep{hu2003hydrodynamics}, or leaf beetles, which harness asymmetric meniscus deformations to generate propulsion \citep{hu2005meniscus}.

Objects floating under capillary forces also interact through the surface deformations they induce \citep{kralchevsky2000capillary}. The flotation force, commonly referred to as the {Cheerios effect} \citep{vella2005cheerios}, represents the most familiar form of capillary interaction. Depending on the meniscus curvature, this interaction can be either attractive or repulsive, and it has been observed for small bubbles \citep{nicolson1949interaction} as well as for small solid objects \citep{dushkin1995lateral}, and may lead to the formation of clusters or rafts \citep{protiere_particle_2023}.

Capillarity also affects the motion of small floaters in waves. \cite{falkovich2005floater} investigated the interaction between a standing wave and small floating particles with varying wetting properties. In this configuration, capillarity induces a weak mean force that drives the particles toward preferential locations, either the nodes or antinodes of the wave, depending on the curvature of the meniscus.

\begin{figure}[tb]
    \centering
    \includegraphics[width=\linewidth]{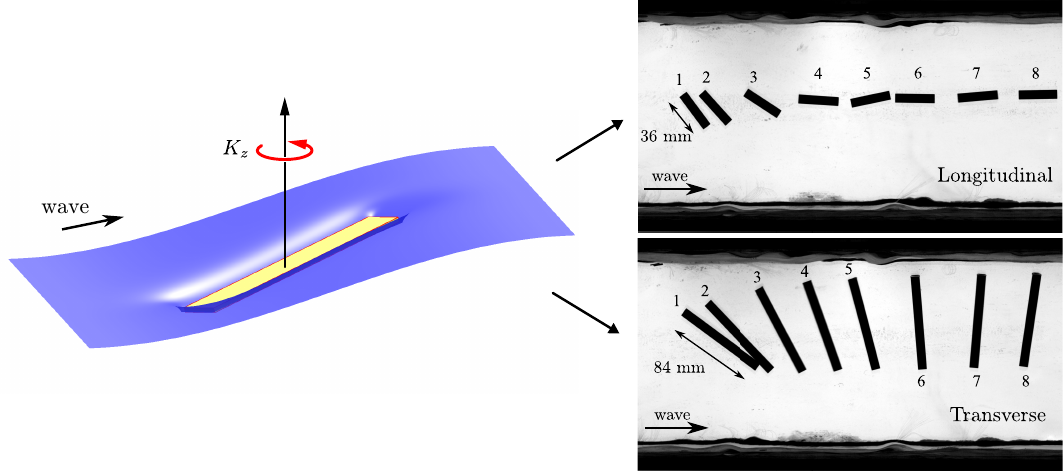}
    \caption{Thin floating plates denser than water placed in a surface wave drift towards a preferential orientation, either longitudinal (for shorter floaters) or transverse (for longer floaters). We study how this preferential orientation phenomenon is affected by capillarity. }
    \label{Fig2_chronophoto}
\end{figure}

In this article, we describe another case of wave-induced mean motion influenced by capillarity. In our experiments, we study thin metallic plates, denser than water, floating at the water surface and drifting within a propagating gravity wave. Viewed from above, the plates exhibit a slow drift in both position and orientation  (see figure~\ref{Fig2_chronophoto}). The angular drift leads to a preferential state of orientation that can be parallel to the direction of wave propagation, called \textit{longitudinal}, or parallel to the wave crests, called \textit{transverse}. While preferential orientations can also arise in the absence of capillary effects \citep{newman1967drift, herreman2024, newman2026drift}, capillary forces are significant for small floaters. The aim of this article is to elucidate the role of capillarity in this problem. 

Capturing capillary forces and moments in wave-floater interaction models is not a simple exercise. One needs to determine the dynamics of a meniscus attached to an object moving in a non-quiescent liquid and then evaluate a contour integral over the instantaneous position of the contact line. In this article, we show that the calculation of the instantaneous capillary forces and moments on the object can be significantly simplified by using a quasi-static extension of  the generalised Archimedes principle of \cite{mansfield1997equilibrium} and \cite{keller1998surface}. For capillary floaters at equilibrium, they showed that the object's weight is balanced with the weight of the fluid displaced by the object and by its meniscus. In this article, we show that, under the Froude-Krylov approximation, ignoring wave diffraction on the object, we can extend this generalised Archimedes principle to dynamic, wave-floater interaction problems.  The sum of the capillary and pressure forces on a small floater in a long wave is still a generalised Archimedes buoyancy force, but it opposes the local Lagrangian acceleration $- \dot{\boldsymbol{u}} -  g  \boldsymbol{e}_z $ of the small volume of fluid displaced by the floater and its meniscus. This quasi-static extension of the generalised Archimedes principle provides a convenient framework to model the motion of arbitrary small floaters in long, slowly evolving waves. 

Let us recall our main previous results on preferential orientation of slender floaters in gravity waves, in the absence of capillarity. In \cite{herreman2024, dhote2025flexible, herreman2026elastic}, we have  considered solid, perfectly flexible and elastic floaters in waves, respectively. For floaters that are short with respect to the wavelength, we found that preferential orientation is governed by a single non dimensional number $F$ and how it compares to a critical value 

\begin{equation}
    F = \frac{k L_x^2}{\hb}, \quad \left \{ \begin{array}{rcl} 
    F < F_c &, & \text{longitudinal} \\
    F > F_c &, & \text{transverse} 
    \end{array} \right.   \label{Fcrit}
\end{equation}
where $k$ is the wavenumber, $L_x$ the floater's length and $\bar{h}$ is the equilibrium immersion. For a homogenous rectangular parallelepiped floater with density ratio $\beta = \rho_p / \rho$ ($\rho_p$ and $\rho$ being the plate and fluid densities) and thickness $L_z$, the equilibrium immersion is $\bar{h} = \beta L_z$ according to the Archimedes principle. The key ingredient explaining the transition from longitudinal to transverse orientation is the spatial variation of submersion depth along the floater's length.

The critical value $F_c$ in Eq.~\eqref{Fcrit} depends on both the shape and the rigidity of the floater. For solid rectangular parallelepipeds, we found $F_c = 60$, in good agreement with experiments \citep{herreman2024}. For perfectly flexible floaters, the submersion depth is constant, resulting in a systematic longitudinal orientation (i.e., $F_c \rightarrow \infty$), here again in agreement with experiments \citep{dhote2025flexible}. For the intermediate case of thin elastic plates with finite bending rigidity $D$, we  found  that \citep{herreman2026elastic}

\begin{equation} \label{eq:Fcdef}
F_c \approx 60  + \frac{5}{42}  \frac{\rho g L_x^4}{D},
\end{equation}
with $g$ the gravitational acceleration. So far, this prediction has not received experimental confirmation.

In this paper, we extend our previous theoretical results by incorporating capillary effects, which are relevant for floaters denser than water, and validate these predictions through experiments performed on centimeter-scale rigid or flexible metal plates (figure~\ref{Fig2_chronophoto}). Applying our quasi-static extension of the generalized Archimedes principle \citep{mansfield1997equilibrium, keller1998surface} leads to a remarkably simple result: Our prediction (\ref{Fcrit})-(\ref{eq:Fcdef}) for preferential orientation remains valid, provided that the liquid density $\rho$ is replaced by an effective density

\begin{equation} \label{eq_def_rhohat}
\hat{\rho} = \rho \left (1 + \frac{2 \lc}{L_y} \right ) ,
\end{equation}
where $\lc$ is the capillary length and $L_y$ is the width of the plate. This influences both the equilibrium submersion $\bar{h} = \hat{\beta} L_z$ (with $\hat{\beta} = \rho_p / \hat{\rho}$), which enters the definition of the number $F$ (\ref{Fcrit}), and the factor $\hat{\rho} g L_x^4 / D$ in the definition of $F_c$ in Eq.~\eqref{eq:Fcdef}. This generalised criterion allows us to account for all of our experimental observations.

The article is structured as follows. We first discuss the static equilibrium of thin plates denser than water in \S \ref{Static}. Experiments on preferential orientation are presented in  \S \ref{PrefOrientation_Expe}.  In \S \ref{KMgen}, we formulate the quasi-static extension of the generalised Archimedes principle. Finally, in \S~\ref{theory}, we incorporate capillary effects into the calculation of the second-order mean yaw moment acting on the thin plate, leading to the generalized criterion (\ref{eq:Fcdef})–(\ref{eq_def_rhohat}). 

\section{Equilibrium submersion of thin metal plates}
\label{Static}

We first consider the floating equilibrium of slender, rectangular and thin plates denser than water. We calculate the theoretical equilibrium submersion depth and compare to experiments.

\subsection{Set-up and notations}

We consider thin rectangular plates of dimensions $L_x, L_y, L_z$ and density  $\rho_p$, with density ratio $\beta = \rho_p / \rho >1$ ($\rho$ is the water density). Given their small thickness, these thin plates can deform elastically. Their deformation  is governed by the gravito-flexural length $L_D$, defined as
\begin{equation} \label{eq_def_Ld}
L_D = \left(\frac{D}{\rho g}\right)^{1/4}, \quad \text{with} \quad D = \frac{EL_z^3}{12(1 - \nu^2)}
\end{equation}
the bending modulus, $E$ the Young Modulus, and $\nu$ the Poisson ratio. We consider two types of floaters:
\begin{subequations}
\begin{eqnarray} \label{ineq_floater_dimensions}
& & L_D \gg L_x \gg L_y \gg L_z : \qquad \mbox{rigid floater} \\
& & L_x \gg L_D \gg L_y \gg L_z : \qquad \mbox{deformable floater}.
\end{eqnarray}
\end{subequations}
Since $L_D\gg L_y, L_z$, we can ignore elastic deformation along the short and middle axis. We also assume that the capillary length $\lc  = ( \gamma/\rho g )^{1/2}$  (with $\gamma$ the surface tension) remains small with respect to the floater's length and to the flexural length, 
\begin{equation}
L_D \gg \lc   \quad , \quad L_x \gg \lc .
\end{equation} 
These inequalities allow us to ignore tension in the plate and to use a simplified (exponential) meniscus profile on the long axis of the floater. 

The thin plates that we use in our experiments are cut from brass sheets, with density ratio $\beta = 8.4$. We consider four different thicknesses $L_z = 0.05, 0.1, 0.15$ and $0.3$~mm, the length $L_x$ varies in the range $20$–$100$~mm, and the width $L_y$ between $1$ and $18$~mm. The plates were laser-cut to ensure clean, defect-free edges and to minimize the intrinsic bending that can occur when using scissor cutting (the influence of intrinsic bending on the equilibrium of the floaters is discussed in section~\ref{sec_natcurv}). 
The bending rigidity $D$ was measured by clamping one extremity of the plate and fitting the resulting deflected shape of the plate. From these measurements, we calculate the flexural lengths in water $L_D = 1.9 \pm 0.1$ and $3.2\pm0.2$ cm for the thicknesses $L_z=0.05$ and $0.1$ mm, in good agreement with the Young modulus of brass, $E\simeq 115 \pm 15$ GPa. For the two other thicknesses, $L_z = 0.15$ and $0.3$ mm, the plates do not show significant deformation with this method. Computing their flexural lengths using equation \eqref{eq_def_Ld} gives $L_D \approx 4.4$ and $7.4$~cm, indicating that these plates can be regarded as essentially rigid for the lengths considered in our experiments.

\subsection{Theoretical submersion depth}\label{Static_model}

\begin{figure}[tb]
    \centering
    \includegraphics[width =\linewidth]{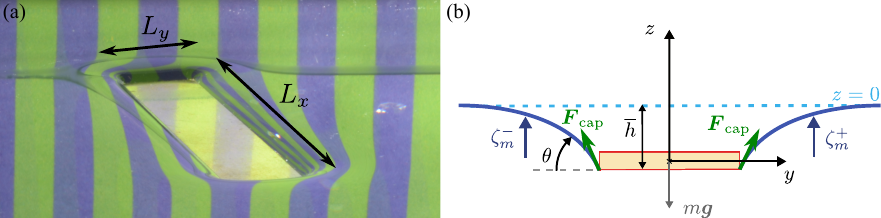}
    \caption{(a) A thin brass rectangular plate of dimensions $L_x = 7$ cm, $L_y = 1$ cm and $L_z = 0.3$ mm floating at the air-water interface. The surface deformation induced by the meniscus is visible through the reflection of a patterned lighting. (b) Force balance on a floating plate. The meniscus $\zeta_m$ along the long edges is pinned at the bottom face of the floater, at $z = -\hb$.}
    \label{Fig1_Photo_ExpStatique}
\end{figure}

When gently placed on the surface of water,  thin metal plates can remain afloat with a meniscus pinned at the edge, as shown by figure~\ref{Fig1_Photo_ExpStatique}(a).   
We define the equilibrium submersion depth $\bar{h}$ as the distance between the bottom face and the undisturbed free surface (figure~\ref{Fig1_Photo_ExpStatique}b). To calculate $\overline{h}$, we express the force balance that includes the weight, buoyancy, the capillary force \citep{vella2006equilibrium,vella2015floating}. Assuming that the contact line is pinned at the lower edge, the vertical force balance writes

\begin{eqnarray}\label{fb0}
 - \rho_p g L_x L_y L_z + \rho g L_x L_y \hb + F_{\text{cap},z} = 0,
\end{eqnarray}
where $F_{\text{cap},z}$ is the vertical component of the capillary force.  For an elongated floater satisfying $L_x \gg L_y$, we can neglect the menisci on the thin $y$-edges, and the capillary force reduces to $F_{\text{cap},z} = 2 \gamma L_x \sin \theta$, with $\theta$ the angle of the meniscus with respect to the horizontal (see figure~\ref{Fig1_Photo_ExpStatique}b). To find $\theta$, we need to solve for the meniscus profile.  From the hydrostatic balance and the Young-Laplace pressure jump at the interface, we obtain $\rho g \zeta_m = \gamma \kappa$, with $\zeta_m$ the meniscus height and $\kappa$ its curvature. Assuming a weak interface slope, the curvature reduces to $\kappa \simeq - \partial^2 \zeta_m / \partial y^2$, and with the boundary conditions $\zeta_m \rightarrow 0$ far from the floater ($|y| \gg L_y/2$) and $\zeta_m = -\hb$ at the lower edge of the floater ($|y| = L_y/2$), we obtain the meniscus profiles on the long edges 

\begin{equation}\label{zetam_eq}
    \zeta_m = -\hb \, e^{-(|y| - L_y/2)/\lc} \quad \text{for} \quad |y| > L_y/2.
\end{equation}
We deduce that $\theta \simeq \sin\theta \simeq \tan\theta\simeq \left.{\partial \zeta_m / \partial y }\right|_{|y| =L_y/2}  \approx \hb / \lc$, so the capillary force is

\begin{subequations} \label{F_cap_z}
\begin{equation} 
    F_{\text{cap},z}   =  2\gamma L_x \hb/\lc.
\end{equation}
Note that this force can also be written as the weight of the water contained within the menisci \citep{mansfield1997equilibrium,keller1998surface},

\begin{equation} 
    F_{\text{cap},z}  =  2 \rho g  \int_{L_y/2}^{+\infty}  L_x \hb \, e^{-(|y| - L_y/2)/\lc} dy = 2 \rho g L_x\lc \hb  .
\end{equation}
\end{subequations}
Using  this expression in the force balance \eqref{fb0}, we obtain  the submersion depth 

\begin{equation}\label{hb}
\hb_\downarrow = \frac{\beta L_z}{1 + 2\lc/L_y}.
\end{equation}
This equation applies to the case of a meniscus pinned on the lower edge. For a pinning on the upper edge, a similar analysis yields

\begin{equation}\label{hbu}
\hb_\uparrow = L_z + \frac{(\beta - 1)L_z}{1 + 2\lc/L_y}.
\end{equation}

\begin{figure}
    \centering
    \includegraphics[width = \linewidth]{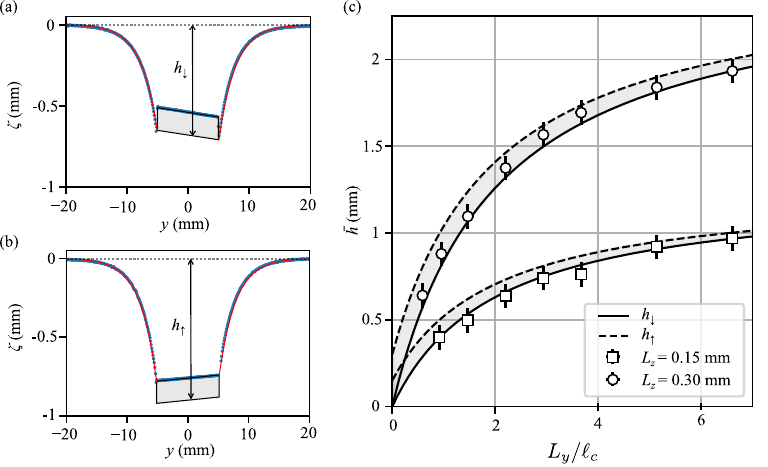}
    \caption{(a,b) Surface height profile across a capillary floater of width $L_y=10$~mm, length $L_x = 60$~mm and thickness $L_z = 0.15$~mm, with meniscus pinned on the lower (a) and upper (b) edge. Note the strongly stretched vertical scale; the residual tilt angle is less than 0.5$^\circ$. (c) Immersion depth $\hb$ (measured at $y=0$) as a function of the normalised width $L_y/\lc$ for two thicknesses $L_z$. The predictions correspond to pinning on the lower edge (full line) and upper edge (dashed line).}
    \label{fig2:StaticImmersion}
\end{figure}

\subsection{Immersion depths measurements}

We first test here the predictions \eqref{zetam_eq}, \eqref{hb} and \eqref{hbu} for our thin brass floaters. We consider for the moment only the thicker plates ($L_z = 0.15$ and 0.3~mm), which remain flat. We carefully place a floater at the surface of distilled water in a Petri dish. The surface profile along $y$ (the middle axis of the plate) is measured at the middle of the floater, using a chromatic confocal sensor (CCS) mounted on a motorized horizontal translation stage, with horizontal and vertical resolution of $10~\mu$m. 
Figures~\ref{fig2:StaticImmersion}(a,b) present two measurements of the surface height profile for the same plate (dimensions $L_x = 60$~mm, $L_y = 10$~mm, $L_z = 0.15$~mm), obtained immediately after deposition at the interface (a), and after a slight perturbation (b). Since brass is hydrophilic (the Young angle is approximately $70^\circ$), pinning of the contact line at the upper edge is, in principle, more stable. However, depending on the cleanliness of the water and on how the floater is deposited, the meniscus may be pinned either at the upper or at the lower edge. Visual inspection alone does not allow to determine whether the contact line is pinned at the upper or lower edge, but the CCS measurements demonstrate that the pinning location can be shifted from the lower edge to the upper edge. Note that in both cases the plates are not perfectly horizontal, probably due to small defects in the pinned contact lines; a residual tilt of approximately $0.5^\circ$ is typically observed (strongly amplified in the figures because of the stretched vertical scale). Mixed configurations, with an upper pinning on one side and a lower pinning on the other side, may also be observed, but are not considered here.

A fit of the menisci with the exponential law \eqref{zetam_eq} with $\ell_c \simeq 2.7$~mm yields excellent agreement with the data, for both pinning at the upper or lower edge. This confirms that the quasi-one-dimensional linear approximation accurately describes the meniscus shape in this parameter range. We repeated the measurements for various plate widths $L_y$, and for the two thicknesses $L_z = 0.15$ and 0.3~mm. The measured immersion depth $\hb$ is shown in figure~\ref{fig2:StaticImmersion}(c) as a function of the normalized width $L_y/\ell_c$. In this range of parameters, the immersion depth remains below $2$~mm (maximum angle $\theta \simeq 35^\circ$), for which the linear approximation remains valid. For each thickness, we also plot the predictions corresponding to a meniscus pinned at the lower edge (Eq.~\ref{hb}) and at the upper edge (Eq.~\ref{hbu}), providing an uncertainty range for the immersion depth. We note that the majority of the data lie slightly closer to the lower-edge prediction~\eqref{hb}. In the following, we therefore assume pinning at the lower edge, which simplifies the analysis. We note, however, that pinning at the upper edge may also occur, especially under the less controlled conditions of the wave-tank experiments, thereby introducing an uncertainty of order $10\%$ in the immersion depth.

\section{Preferential orientation in waves: experiments}\label{PrefOrientation_Expe}

Left adrift in gravity waves, we observe that slender thin plates slowly rotate towards a preferential orientation, as illustrated in the chronophotographies of figure~\ref{Fig2_chronophoto}. We conducted an extensive series of experiments to systematically characterize this preferential orientation for a range of wavelengths and plate dimensions, including length, width, and thickness.

\subsection{Experimental setup}

\begin{figure}[bt]
    \centering
    \includegraphics[width = \linewidth]{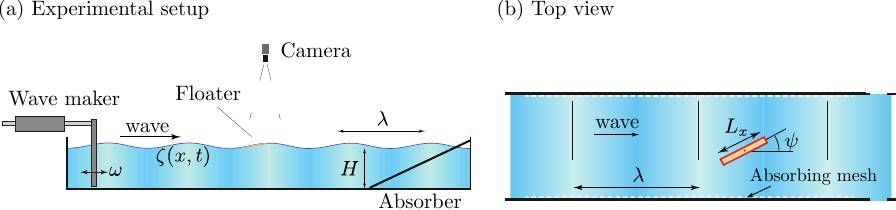}
    \caption{Experimental setup. (a) Side view; a 4 m long, 18 cm large wave tank is filled at height $H = 22$ cm. A wave maker oscillates at frequency $\omega$, generating waves of wavelength $\lambda$. (b) Top view of a floater drifting in a propagating wave. The yaw angle $\psi$ is the angle between the floater long axis and the direction of wave propagation. A nylon mesh is attached to the walls to absorb parasitic surface waves.}
    \label{Fig3_Setup&Cuve}
\end{figure}

Experiments are performed in a  rectangular wave channel, $4$ m long, $18$ cm wide, filled with water at height $H = 22$ cm (see figure~\ref{Fig3_Setup&Cuve}(a)). {The lateral walls are covered with a wet nylon mesh fabric of mesh size $0.1$~mm, designed to mitigate parasitic capillary ripples induced by the periodic wetting and dewetting of the glass by the passing wave crests and troughs \citep{kim2016,monsalve2022space}.} The waves are generated by a piston driven by a linear motor, oscillating in a sinusoidal motion. The waves are attenuated at the end of the tank by a sloping plate of length 1~m and angle $12^\circ$. The wave frequency $f = \omega/2\pi$ is varied between $1.5$ and $2.5$ Hz, corresponding to wavelengths in the range $\lambda \simeq 26 - 66$~cm. Since $\lambda \gg\ell_c$, we only consider surface waves in the gravity regime. The characteristic response time of the meniscus can be estimated as

\begin{equation} \label{response_time}
\tau_{m} = \left(\frac{\gamma}{\rho g^3}\right)^{1/4} \approx 0.01~\text{s}
\end{equation}
which is very short compared with the wave period. Accordingly, the meniscus can be assumed to adjust quasi-statically to the instantaneous position of the floater and the wave.

For each wavemaker frequency, the wave amplitude $a$ and wavelength $\lambda$ are measured using a camera located on the side of the wave tank, with a precision of  $1$ mm and $1$ cm, respectively. Because of imperfect wave attenuation at the sloping plate, the wave contains a small steady component, resulting in temporal oscillations of the amplitude of the order of 5\%. The wave amplitude is adjusted in the range $a \simeq 6 - 16$ mm to maintain a nearly constant slope $\epsilon = ak \simeq 0.14 \pm 0.01$. This value is selected to achieve a rapid drift and reorientation, while ensuring that the waves remain within the linear regime and that the floaters stay afloat.

Each experiment was conducted as follows. A floater is gently deposited at the water surface, with particular care taken to prevent its tips from piercing the surface. The floater is initially oriented at an angle close to $\psi \simeq 45^\circ$. The wave maker is then activated, and the floater motion is recorded using a camera positioned above the tank. For each combination of floater length and width, the experiment was repeated several times; only trajectories for which the floater remained afloat and approximately centred in the channel were retained, to avoid interactions with the lateral walls. Between runs, a waiting period of approximately $10$ minutes was allowed for waves and residual currents to dissipate.

\subsection{Rigid capillary floaters}\label{Exp_rigids}

\begin{figure}[tb]
    \centering
    \includegraphics[width = \linewidth]{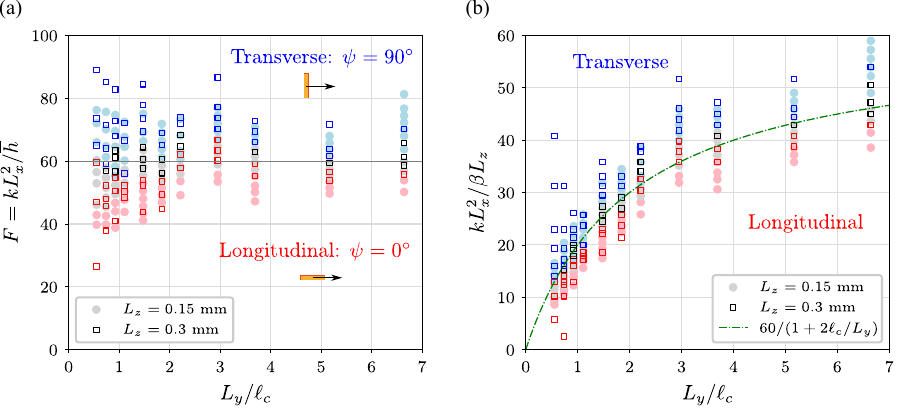}
    \caption{Preferential orientation of rigid capillary floaters in the plan (a) ($F = kL_x^2/\hb, L_y/\lc$) and (b) ($kL_x^2/\beta L_z,L_y/\lc$). Red (resp. blue) points are floaters that rotate towards the longitudinal (resp. transverse) orientation. Circles in pale colours correspond to $L_z = 0.15$ mm, and squares in plain colours to $L_z = 0.3$ mm. Gray and black symbols denote floaters for which no preferential orientation can be identified.}
    \label{Fig4_Exp_F_vs_Lylc}
\end{figure}

We first focus on rigid floaters, of thickness $L_z  = 0.15$ and 0.3~mm, for which $L_x \gg L_D$. Figure~\ref{Fig4_Exp_F_vs_Lylc}(a) shows the preferential orientation of the floater in the $(F, L_y/\ell_c)$ parameter space, where $F = kL_x^2/\bar{h}$ and $\bar{h} = \beta L_z/(1 + 2\lc/L_y)$. The diagram reveals a clear transition: floaters with $F > F_c \simeq 60 \pm 10$ rotate toward the transverse orientation ($\psi  \rightarrow 90^\circ$, blue symbols), while those with $F < F_c$ rotate toward the longitudinal one ($\psi \rightarrow 0^\circ$, red symbols).  Gray symbols denote cases for which no well-defined preferential orientation can be identified, either because the rotation is too slow or because multiple final orientations are observed.

The observed transition  is in excellent agreement with the theoretical value $F_c = 60$ derived in \cite{herreman2024} for non-capillary floaters. This indicates that this criterion also applies for capillary floaters, provided that we use in the definition of $F$ the submersion depth $\bar{h} =\beta L_z/(1 + 2\lc/L_y)$ that incorporates capillary effects. To better asses the effect of capillarity, we plot  in figure~\ref{Fig4_Exp_F_vs_Lylc}(b) the same data using the original definition of $F$ with $\hb = \beta L_z$.  With this alternate definition, the transition between longitudinal and transverse orientations is no longer characterized by a constant threshold, but instead follows the curve given by $kL_x^2 / \beta L_z = 60/(1+2\lc/L_y)$. This representation also suggests that very thin needle-like floaters with $L_y/\lc \rightarrow 0$ will mainly rotate towards the transverse orientation.

\subsection{Elastic capillary floaters}\label{elastocapfloaters_exp}

We now consider the thinner plates, of thickness $L_z = 0.05$ and 0.1~mm, for which elastic deformations can no longer be neglected. For these experiments, the floater width is fixed at $L_y = 1$~cm, and only the length $L_x$ is varied.

\begin{figure}[tb]
    \centering
    \includegraphics[width=.75\linewidth]{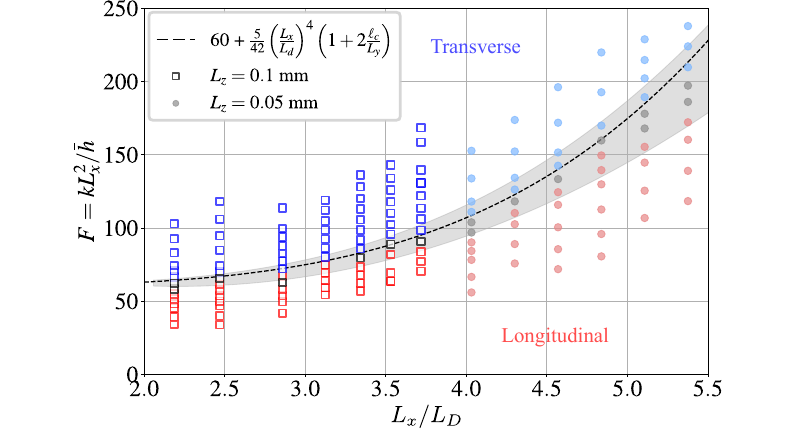}
    \caption{Preferential orientation of elasto-capillary floaters in the ($F,L_x/L_D$) plane, for two floater thickness $L_z$. Red: Longitudinal orientation; blue: transverse orientation; black: uncertain orientation. The gray area represents the uncertainty region. The black dashed line shows the predicted transition $F_c$ (equation \ref{eq:Fcdef}).}
    \label{F_vs_LxlD_ManipsModel}
\end{figure}

The preferential orientation for these floaters is shown in figure~\ref{F_vs_LxlD_ManipsModel} in the parameter space $(F = k L_x^2 / \hb, L_x/L_D)$. The data is represented using the same colour scheme as in figure~\ref{Fig4_Exp_F_vs_Lylc}: blue symbols for transverse floaters, red for longitudinal floaters and gray for undefined orientation. We obtain an excellent agreement with the elasto-capillary prediction \eqref{eq:Fcdef} using the effective density $\hat \rho$ \eqref{eq_def_rhohat}. This shows that the combined effects of rigidity and capillarity, through the parameters $\hat \rho$ and  $L_x/L_D$, have a strong influence on the transition between longitudinal and transverse orientation. 

\section{Forces and moments on small floaters in long waves}
\label{KMgen}

We aim to model the motion of a small floating object together with its surrounding meniscus in long-wavelength, inviscid gravity waves. To do so, it is necessary to evaluate the instantaneous pressure and capillary forces and moments acting on the floater.

In this section, we show that, under the diffractionless, Froude-Krylov assumption, all forces and moment can be calculated as volume integrals over the virtual volume of the displaced fluid, including the meniscus. The mathematical operations that lead to this formulation are similar to those that lead to the generalised Archimedes principle in equilibrium conditions \citep{mansfield1997equilibrium,keller1998surface}. This quasi-static extension   can be useful in other floater-wave interaction problems with significant capillary effects, provided that the floater length is much smaller than the wavelength.

\subsection{Pressure formulation of capillary forces}

\begin{figure}[tb]
    \centering
    \includegraphics[width=0.99\linewidth]{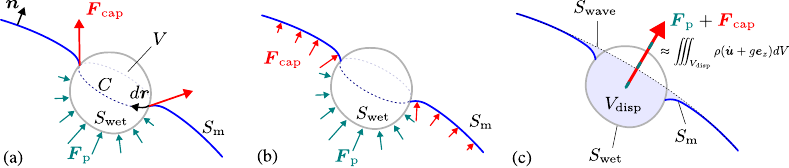}
    \caption{Pressure and capillary forces on an arbitrary floating body of volume $V$ surrounded by a meniscus. (a) The pressure force $\boldsymbol{F}_\text{p}$ acts on the wetted surface $\Ssub$, and the capillary force  $\boldsymbol{F}_\text{cap}$ acts on the contact line $C$. (b) The capillary force is identical to the pressure force on the free surface $S_\text{m}$. (c) In the Froude-Krylov approximation, the resultant force $\boldsymbol{F}_\text{p} + \boldsymbol{F}_\text{cap} $ acts as a generalized buoyancy force opposing the local acceleration $-\dot{\boldsymbol{u}} - g \boldsymbol{e}_z$ of the displaced volume of liquid $V_\text{disp}$ enclosed by the wetted surface $\Ssub$, the meniscus $S_\text{m}$ and the unperturbed wave surface $S_\text{wave}$.}
    \label{Fig_gene_surfaces}
\end{figure}

We consider a floater of arbitrary shape at some instantaneous position in a wave (figure  \ref{Fig_gene_surfaces}a). 
We show that the resultant of the pressure and capillary forces on this floater can be calculated as a pressure force on the surface of the object and on its meniscus. In our case the meniscus is pinned but our approach holds with a moving contact line, whose angle is close to the value fixed by the Young-Dupré law in the quasi-static case.

We denote by $V$ the floater volume, $\Ssub$ its wetted surface, delimited by the contact line $C$. The free surface is denoted $S_{\text{m}}$, and $\bs{n}$ is the outward unit normal. The surface element $d \bs{S}$ is everywhere defined away from the liquid. We denote $d \br$ as the line element along the contact line, oriented according to the right-hand rule with respect to the free-surface normal $\bs{n}$.

We note $p$ the total instantaneous pressure in the water, $p_0$ the atmospheric pressure, $\br$ the position vector, and $\brc$ the centre of mass of the floater. The pressure force $\boldsymbol{F}_\text{p}$ and the capillary force  $\boldsymbol{F}_\text{cap}$ are given by 

\begin{equation}
\boldsymbol{F}_\text{p} =
 \iint_{\Ssub} (p-p_0) \bdS   , \qquad 
 \boldsymbol{F}_\text{cap}=\oint_{C} \gamma \bs{n} \times d \br .
\end{equation}

Using Stokes theorem, the contour integral can be written as a surface integral of the curvature $\kappa = \boldsymbol{\nabla} \cdot \boldsymbol{n}$ along the contour. Then, using the Young-Laplace boundary condition, $p - p_0 = \gamma  \kappa $, we can rewrite this integral as

\begin{equation}
 \boldsymbol{F}_\text{cap} =\iint_{S_\text{m}} \gamma  \kappa  \,\bdS  =   \iint_{S_\text{m}}   (p-p_0) \, \bdS .
\end{equation}
The sum of the pressure and capillary force can therefore be expressed as a surface integral of pressure over the total surface $\Ssub \cup S_\text{m}$:

\begin{equation}
\boldsymbol{F}_\text{p}  + \boldsymbol{F}_\text{cap}= 
 \iint_{\Ssub \cup S_\text{m} } (p-p_0) \bdS.   \label{eqFexact}
 \end{equation}
 For the pressure and capillary moments $\boldsymbol{K}_p$ and  $\boldsymbol{K}_c$ 
 
\begin{equation}
\boldsymbol{K}_\text{p} = \iint_{\Ssub} (\br - \brc)\times  (p-p_0) \bdS      \quad , \quad \boldsymbol{K}_\text{cap} =  \oint_{C} (\br - \brc)\times ( \gamma \bs{n} \times d \br  )  
\end{equation}
we find in a similar way that their sum reduces to 

 \begin{equation}  \label{eqKexact}
\boldsymbol{K}_\text{p} + \boldsymbol{K}_\text{cap} =
 \iint_{\Ssub \cup S_\text{m} }  (\br - \brc)\times  (p-p_0) \bdS.
\end{equation}
Physically, equations~\eqref{eqFexact} and \eqref{eqKexact} indicate that the liquid acts on the object by pushing both on the wetted surface and on the meniscus. This formulation also applies in the presence of spatially varying surface tension (Marangoni effect), and may also be extended to liquids with arbitrary stress tensors $\boldsymbol{\sigma}$, by replacing $ (p-p_0) \bdS$ with $(-\boldsymbol{\sigma} -p_0 \boldsymbol{1}) \cdot \bdS$. 



\subsection{Forces on small objects in long waves in the Froude-Krylov approximation}

In the particular situation of a small floating object in a long wave, we reduce the expression of the resultant capillary and pressure forces to a simple volume formulation. At the scale of the object, both the pressure and velocity fields vary weakly in space. This implies that, at leading order, the floater is primarily advected by the local, nearly homogeneous flow, such that the instantaneous velocity difference between the object and the surrounding liquid remains small. The floater therefore acts as a negligible disturbance to the long wave, allowing diffraction corrections to be neglected.

Within the Froude–Krylov approximation, the pressure $p$ in Eqs.~\eqref{eqFexact} and \eqref{eqKexact} is approximated to that of the incident wave alone. Since we have by definition  $p=p_0$ on the free surface of the unperturbed gravity wave  $S_\text{wave}$,  we can add the zero contribution, $\iint_{S_\text{wave} }  (p-p_0) \bdS  = \boldsymbol{0} $,  to the integral \eqref{eqFexact}. This yields an integral over the closed surface $\Sigma = \Ssub \cup S_\text{m} \cup S_\text{wave}$, which encloses the entire displaced fluid volume $V_{\text{disp}}$, i.e., the volume displaced by the floater and its meniscus, relative to the reference situation with wave and no floater. By orienting $ \bdS$ towards the interior of this volume, we can write the resultant pressure and capillary force as:

\begin{equation}
\boldsymbol{F}_{p} + \boldsymbol{F}_{c} = 
 \oiint_{\Sigma  } (p-p_0) \bdS ,
\end{equation} 
where the pressure difference $p-p_0$ is defined throughout the displaced volume $V_{\text{disp}}$ enclosed by $\Sigma$. 
We apply the divergence theorem to this integral, with a negative sign to take into account that $ \bdS$ is oriented towards the interior. Using then Euler's equation, $ \rho ( \partial_t \boldsymbol{u}  + ( \boldsymbol{u}  \cdot  \boldsymbol{\nabla} ) \boldsymbol{u}  ) = -  \boldsymbol{\nabla}p -  g  \boldsymbol{e}_z  $, we obtain

\begin{equation} \label{Fvol}
\boldsymbol{F}_{p}+ \boldsymbol{F}_{c} = 
 \iiint_{V_{\text{disp}}  } ( -\boldsymbol{\nabla}p ) d V    =  \iiint_{V_{\text{disp}}  } \rho ( \partial_t \boldsymbol{u}  + ( \boldsymbol{u}  \cdot  \boldsymbol{\nabla} ) \boldsymbol{u}   +  g  \boldsymbol{e}_z  )  d V .
\end{equation} 
For the sum of the pressure and capillary moment, similar reductions yield 

 \begin{equation}  \label{Kvol}
\boldsymbol{K}_{p}  + \boldsymbol{K}_{c}=
 \iiint_{V_{\text{disp}}  } (\br - \brc)\times   \rho ( \partial_t \boldsymbol{u}  + ( \boldsymbol{u}  \cdot  \boldsymbol{\nabla} ) \boldsymbol{u}   +  g  \boldsymbol{e}_z  )  d V .
\end{equation} 
In the absence of flow ($\boldsymbol{u} =\boldsymbol{0}$), we recover the generalised Archimedes buoyancy force of \cite{mansfield1997equilibrium} and \cite{keller1998surface}: the sum of the capillary and pressure force equals the weight of the fluid displaced by the object and its meniscus. We can give the following physical interpretation of these equations: the resultant capillary and pressure forces on a small floater in a long wave reduce to an Archimedes-like buoyancy force opposing the effective acceleration  $-   \dot{\boldsymbol{u}} -  g  \boldsymbol{e}_z $ (where the dot denotes the Lagrangian derivative) of the displaced volume of liquid. Note that, by adopting the Froude-Krylov approach, we implicitly assume that the meniscus adapts quasi-statically to the floater and surface motion, without emitting capillary waves. Hence equations \eqref{Fvol} and \eqref{Kvol} provide a quasi-static extension of \cite{mansfield1997equilibrium} and \cite{keller1998surface} to the case of time-dependent flow.


\section{Preferential orientation in waves: theory}
\label{theory}



We derive in this section the theoretical model for the preferential orientation of a thin capillary floater drifting in a propagating gravity wave. We follow the methodology of  \cite{herreman2026elastic}, in which the floater is modeled as a thin, deformable plate.

\subsection{Problem definition}

\begin{figure}[bt]
    \centering
    \includegraphics[width = \linewidth]{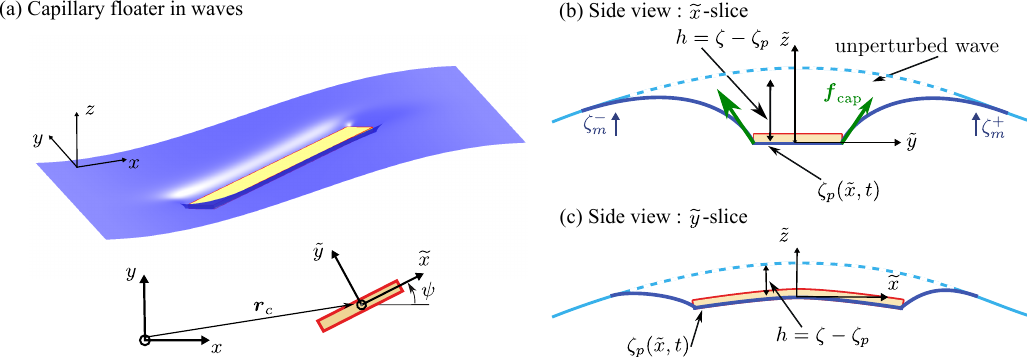}
    \caption{(a) Elastic plate with menisci pinned on its bottom face floating in a surface wave. Two side views show (b) the pinned menisci $\zeta_m^\pm$ along the short dimension $\yt$ and (c) the plate deformation $\zeta_p(\xt,t)$ along the long dimension $\xt$. The light-blue dashed line is the wave in the absence of object, and we show the local generalised immersion height $h = \zeta - \zeta_p$. The menisci at the tips $\xt=\pm L_x/2$ are shown on (a) and (c) but are not taken into account in the model.}
    \label{FIG4_Model}
\end{figure}

The system is sketched in figure~\ref{FIG4_Model}. The incoming wave is idealised as a linear inviscid potential wave in infinitely deep water. In the reference frame $(O,x,y,z)$, the surface elevation and velocity potential are

\begin{equation}
    \zeta (x,t) = a \sin(kx - \omega t), \hspace{.5 cm} {\phi} (x,z,t) = - \frac{a \omega}{k} e^{kz} \cos(kx - \omega t), \label{eqflow}
\end{equation}
where $a, k, g$ denote the wave amplitude, wave number and gravitational acceleration, with the frequency satisfying $\omega = \sqrt{gk}$. The {fluid velocity is $\bs{u} = \nabla \phi$ and the pressure is $p = p_0 - \rho g z - \rho \partial_t\phi$}, where $p_0$ is the atmospheric pressure and $\rho$ the fluid density. We do not need second order corrections to this wave as they do not affect the preferential orientation in our model, as shown in \cite{herreman2024}.

To describe the floater motion, we introduce a second reference frame $(C,\wt{x},\wt{y},\wt{z})$ attached to $C$, the projection of the centre of the strip on the $z=0$ plane. We align the $\wt{x}$-axis with the long axis,  the $\wt{y}$-axis with the middle axis and $\wt{z}=z$ (figure~\ref{FIG4_Model}a). The yaw angle $\psi(t)$ is the angle between the $x$-axis and the $\wt{x}$-axis. Relative to $O$, the floater centre $C$ has coordinates $x_c (t)$ and $y_c$ in the laboratory frame, with $y_c$ an arbitrary constant. The bending of the floater is represented by the function $\zeta_p (\wt{x},t)$ that locates the bottom face of the floater, as shown in figure~\ref{FIG4_Model}(b). We ignore bending deformations along $\wt{y}$ since $L_y \ll L_D$. In the laboratory frame, the points of this bottom face are at 

\begin{eqnarray}
     x \approx x_c (t) + \xt \cos  \psi (t), \quad\quad y  \approx y_c + \xt \sin \psi (t), \quad\quad z  =  \zeta_p (\wt{x},t) + \zt.
\label{tf}
\end{eqnarray}
By varying $\wt{x} \in [-L_x/2,L_x/2]$, $\wt{y} \in [-L_y/2,L_y/2]$,  we cover the entire bottom surface.

%

The shape of the bottom face of the plate $\zeta_p$ is solution of the Kirchhoff-Love equation 
\citep{timoshenko1959plates,landau1986elasticity}: 
\begin{eqnarray}
    D \frac{\partial^4\zeta_p}{\partial\xt^4} - \underbrace{\gamma \frac{\partial^2\zeta_p}{\partial\xt^2}}_{\text{negl}} + \underbrace{\rho_pL_z\frac{\partial^2\zeta_p}{\partial t^2}}_{\text{negl}} = p |_{z=\zeta_p } - p_0 - \rho_p g L_z + f_{\text{cap},z},
    \label{eq_plaque_gen}
\end{eqnarray}
Compared to \cite{herreman2026elastic}, there are two extra capillary terms. The term $f_{\text{cap},z}$ stands for the capillary forces per unit area, due to the meniscus that pulls on the floater on the long edges $\yt = \pm L_y/2$ (figure~\ref{FIG4_Model}b). The term $\gamma \partial^2\zeta_p/\partial\xt^2$ represents the tension in the strip, but for weakly deformable floaters with $L_D > \lc$, it is negligible. The inertial term $\rho_pL_z{\partial^2\zeta_p}/{\partial t^2}$ can also be neglected because the plate is thin and the gravity wave frequency is low. 
In the right hand side, we need to inject the incoming wave pressure and first simplify

\begin{eqnarray}
    p |_{z=\zeta_p }  &=& \nonumber p_0 - \rho g \zeta_p - \rho \partial_t\phi|_{z=\zeta_p},\\
    & \approx &p_0 - \rho g \zeta_p + \rho g(1 + k\zeta_p)\zeta,\nonumber\\
    & \approx & p_0 + \rho g (\zeta - \zeta_p).\label{eq_p_zetap}
\end{eqnarray}
We neglect the term $\rho g k\zeta \zeta_p$ as we only need the pressure at first order in wave amplitude to calculate the first order plate deformation. Injecting \eqref{eq_p_zetap} in \eqref{eq_plaque_gen} yields

\begin{eqnarray}
    D \frac{\partial^4\zeta_p}{\partial\xt^4} + \rho g\zeta_p = \rho g\zeta - \rho_p g L_z + f_{\text{cap},z}.
    \label{eq_plaque_fz_a_detailler}
\end{eqnarray}
We now consider the capillary term $f_{\text{cap},z}$ in equation \eqref{eq_plaque_fz_a_detailler}.  To find this force, we need to calculate the shape of the meniscus $z=\zeta_m$ on  the long edges $\yt = \pm L_y/2$ of the floater. Since $l_c \ll L_x$ and $l_c \ll \lambda$, we expect that the meniscus varies much more rapidly in the $\wt{y}$-direction than in the $\wt{x}$-direction. We also know that the response time of the meniscus is short compared to the wave-period (see Eq. \eqref{response_time}), so that it can be considered as quasi-static: the meniscus immediately adapts to local floater shape at $\yt = \pm L_y/2$  and to the instantaneous wavy surface deformation, far from the floater  $\yt \rightarrow +\infty$.  Both combined, we can calculate the meniscus shape $\zeta_m^{\pm}$ on both sides $\yt = \pm L_y/2$ of the floater, as an exponential solution of the problem

\begin{equation}
\gamma \frac{\partial^2\zeta_m^{\pm}}{\partial\yt^2} - \rho g\zeta_m^{\pm} \approx 0
\end{equation}
with boundary conditions

\begin{equation}
\zeta_m^{\pm} |_{\yt = \pm L_y/2}=  \zeta_p  (\xt,t)  \quad \text{and} \quad \zeta_m^{\pm}  \xrightarrow{\yt \rightarrow +\infty}  \zeta(x,t).
\end{equation}
We suppose here that the menisci are pinned at the lower edge of the floater. This gives the solution

\begin{eqnarray}\label{eq_htm}
    \zeta_m^{\pm}(\xt,\yt,t) = \zeta(x,t) + (\zeta_p(\xt,t) - \zeta(\xt,t))\exp \left(-\frac{\left|\yt\right| - L_y/2}{l_c} \right)
\end{eqnarray}
which generalises equation \eqref{zetam_eq}. We can now write the vertical component of the capillary forces per unit of surface as 

\begin{equation}
f_{\text{cap},z} =\frac{ \gamma}{L_y} \, ( \sin \theta_m^+ +\sin \theta_m^- )
\end{equation}
where $\theta_m^{\pm}(\xt,t)$ is the angle between the meniscus and $\yt$ (figure~\ref{FIG4_Model}). In our linear approach, we always have a small meniscus slope, so  $\theta_m^{\pm} \simeq \partial \zeta_m^\pm/\partial \yt$, yielding

\begin{eqnarray}
f_{\text{cap},z} & = & \frac{\gamma}{L_y} \left.\frac{\partial \zeta_m^{+}}{\partial \yt} \right|_{\yt = L_y/2} - \frac{\gamma}{L_y} \left.\frac{\partial \zeta_m^{-}}{\partial \yt} \right|_{\yt = -L_y/2}=\rho g\frac{2\lc}{L_y}(\zeta - \zeta_p).
\end{eqnarray}
We inject this expression in the plate equation \eqref{eq_plaque_fz_a_detailler} and find

\begin{eqnarray}
    D \frac{\partial^4\zeta_p}{\partial\xt^4}  + \rho \left(1 + \frac{2\lc}{L_y}\right)  g \zeta_p =  \rho  \left(1 + \frac{2\lc}{L_y}\right)  g\zeta - \rho_p g L_z.
    \label{eq_plaque_a_resoudre}
\end{eqnarray}
The boundary conditions at $\xt = \pm L_x/2$ are those of a free plate (no bending moment and no shear force on the end-caps). In our approximation, we have also ignored capillary tension and since $L_x \gg L_y$, we can also ignore the effect of meniscus on the surfaces $\wt{x} = \pm L_x/2$. We then have boundary conditions

\begin{eqnarray}
    \left.\frac{\partial^2\zeta_p}{\partial \xt^2} \right|_{\xt = \pm L_x/2} = 0, \quad \left.\frac{\partial^3\zeta_p}{\partial\xt^3}\right|_{\xt = \pm L_x/2} = 0.\label{BC_zeta_p}
\end{eqnarray}
The problem for the first order plate deformation  $\zeta_p $ is now fully defined and can be solved. From the unperturbed free surface $\zeta$ and $\zeta_p $, we can deduce the local submersion depth as

\begin{equation} \label{immersion}
h  (\wt{x},t) = \zeta (\wt{x},t) - \zeta_p  (\wt{x},t) .
\end{equation}
This quantity plays an important role in our model.

Equations of motion for $x_c(t)$ and $\psi(t)$ are given by Newton's law ($x$-component) and the angular momentum theorem ($z$-component)

\begin{eqnarray}  
        \frac{d}{dt}\underbrace{\left(\int_{V} \rho_p  v_x \, dV\right)}_{ =\, M\dot{x}_c} = F_x, \quad\quad \frac{d}{dt}\underbrace{\left(\int_{V}\rho_p  \, \bs{e}_z \cdot (\bs{r} - \bs{r}_c) \times \bs{v}  \, dV\right)}_{= \, I_{zz}\dot{\psi}} = K_z,
\label{Motion_eq}
\end{eqnarray}
These integrals cover  $V$, the total floater volume. We denote $M=\rho_p  L_x L_y  L_z$ the floater mass and we approximate the moment of inertia of our elongated floater as $I_{zz} \approx M L_x^2 /12$. In our Froude-Krylov model and using the volume formulation of section \S \ref{KMgen}, we have 

\begin{subequations} \label{FetK}
\begin{eqnarray} 
    F_x  & = &  \int_{\Vdisp} \rho ( \partial_t u_x  +  \underbrace{( \boldsymbol{u}  \cdot  \boldsymbol{\nabla} ) u_x}_{\text{negl}} )  dV \label{Fx}\\
     K_z & = &   -\int_{\Vdisp}  (y - y_c)\rho ( \partial_t u_x +  \underbrace{  ( \boldsymbol{u}  \cdot  \boldsymbol{\nabla} ) u_x }_{\text{negl}})    \label{Kz}   dV.
\end{eqnarray}
\label{FxKz_volume}
\end{subequations}
We can ignore the nonlinear term in the calculation of the first and second order force and moment, because the time-average of $\overline{( \boldsymbol{u}  \cdot  \boldsymbol{\nabla} ) u_x} = 0$ vanishes exactly with our gravity wave. With capillarity, the total horizontal force and yaw moment are integrals over the volume of fluid comprised between the bottom plate of the floater, the surface of the meniscus, and the free surface in the absence of object.

For the thin and slender floaters in long gravity waves, we can further simplify these volumes integrals.  First of all, we can use the small floater assumption $\lambda \gg L_x$ and $\lambda \gg l_c$ to approximate

\begin{equation}
 \partial_t u_x  \approx \partial_t u_x  |_{z = \zeta_p} 
\end{equation}
which neglects all the $z$-dependences in the integrand. We then parametrise the displaced volume in two parts. The volume above the bottom face with $\xt \in [-L_x/2,L_x/2]$, $\yt \in \pm [-L_y/2,L_y/2]$ and $z  \in [\zeta_p,\zeta]$ and the volume above the menisci with $\xt \in [-L_x/2,L_x/2]$, $|\yt| \in \pm [L_y/2,L]$ and  $z \in [\zeta_m^{\pm},\zeta]$, where $\pm$ stands for the meniscus pinned at $\yt = \pm L_y/2$. We introduce a cutoff $L$ that describes the meniscus size, such that $\lambda \gg L \gg \lc$. For the force this gives

\begin{eqnarray}
    F_x & \approx & \int_{-L_x/2}^{L_x/2}   \left.\rho \partial_t u_x\right|_{z=\zeta_p}  (\zeta - \zeta_p) L_y   d\xt  \nonumber \\
    & + &   \int_{-L_x/2}^{L_x/2}   \left.\rho \partial_t u_x\right|_{z=\zeta_p}  \left [ \int_{L_y/2}^{L} (\zeta - \zeta_m^+) d\yt  + \int_{-L}^{-L_y/2} (\zeta - \zeta_m^-) d\yt\right] d\xt . \label{Fx_integrated_z} 
\end{eqnarray}
and the expression of the moment is similar. The terms on the second line are due to capillarity and they can further be simplified by injecting the shape of the menisci $\zeta_m^\pm$ given in Eq.~\eqref{eq_htm}. We evaluate these $\yt$-integrals and take the limit $L\rightarrow +\infty$. This leads to the formula

\begin{subequations} \label{FxKz_line}
    \begin{eqnarray} 
        F_x & \approx & \int_{-L_x/2}^{L_x/2}  \rho   \partial_t u_x |_{z=\zeta_p (\wt{x},t)} \,  h (\wt{x},t) \, (L_y  + 2 \lc ) \,  d \wt{x} , \\
        K_z &  \approx  &  -   \int_{-L_x/2}^{L_x/2}  \rho    (y - y_c)  \partial_t u_x |_{z=\zeta_p (\wt{x},t)}  \,  h (\wt{x},t) \, (L_y  + 2 \lc )  \,  d \wt{x}.
    \end{eqnarray}
\end{subequations}
In both line integrals, we see that the spatially varying submersion along the long axis appears through the function $h (\wt{x},t) = \zeta - \zeta_p$. The effect of the capillary forces is the extra $2 l_c $ factors in the integrals. 

\subsection{Effective density analogy} \label{rho_eff}

The model developed in the previous section includes capillarity but it is entirely analogous to the original non-capillary problem considered by \cite{herreman2026elastic}. Indeed, by introducing the effective density

\begin{equation}
\hat{\rho} =  \rho \left(1 + \frac{2\lc}{L_y}\right),
\end{equation}
the plate deformation problem becomes 

\begin{eqnarray}
    D \frac{\partial^4\zeta_p}{\partial\xt^4}  + \hat{\rho} g \zeta_p =  \hat{\rho} g\zeta - \rho_p g L_z,
\end{eqnarray}
which is identical to that in \cite{herreman2026elastic}.  The force and moment integrals are

\begin{subequations} \label{FxKz_line}
    \begin{eqnarray} 
        F_x & \approx & \int_{-L_x/2}^{L_x/2}  \hat{\rho}   \partial_t u_x |_{z=\zeta_p (\wt{x},t)} \,  L_y \, h (\wt{x},t)  \,  d \wt{x} , \\
        K_z &  \approx  &  -   \int_{-L_x/2}^{L_x/2}  \hat{\rho}    (y - y_c)  \partial_t u_x |_{z=\zeta_p (\wt{x},t)}  \,  L_y \,  h (\wt{x},t)    \,  d \wt{x},
    \end{eqnarray}
\end{subequations}
with $h = \zeta - \zeta_p$. The preferential orientation of a capillary floater in a fluid of density $\rho$ is therefore identical to that of a non-capillary (large-scale) floater in a fluid of effective density $\hat{\rho}$. The equilibrium submersion depth indeed writes $\overline{h} = \hat{\beta} L_z$ with  $\hat{\beta} = \rho_p /  \hat{\rho}$, and it is this $\overline{h}$ that enters the definition of the non-dimensional number $F = k L_x^2 / \overline{h}$. Adapting the non-capillary prediction of \cite{herreman2026elastic}, we therefore expect a longitudinal orientation when $F< F_c$ and a transverse orientation when $F > F_c$, with

\begin{equation}\label{Fc_elastocap_guess}
F_c  \approx  60  + \frac{5}{42}  \frac{\hat{\rho} g L_x^4}{D} .
\end{equation}
In the next section, we present a perturbative analysis that confirms this prediction.

\subsection{Preferential orientation in the weakly deformable limit}
\subsubsection{Non-dimensionalisation and multiscale analysis}
We non-dimensionalise space in units of $k^{-1}$, time in units $\omega^{-1}$, and mass in units $M$. We denote 

\begin{equation}\label{Adim}
    \epsilon = ka , \quad l_{x,y,z} = kL_{x,y,z}, \quad \hat{l}_D = k (D/ \hat{\rho} g)^{1/4}
\end{equation}
the wave slope, non-dimensional floater sizes and non-dimensional flexural length. For the incoming wave we have in non-dimensional form 

\begin{subequations}
    \begin{eqnarray}
        \zeta  &=& \epsilon \sin (x_c - t  +  c_\psi  \xt),\\
        \partial_t u_x |_{z= \zeta_p } &=& - \epsilon e^{\zeta_p } \cos{(x_c - t +  c_\psi \xt)},\\
        p|_{z= \zeta_p } - p_0  &= &- \zeta_p + \epsilon e^{\zeta_p } \sin{(x_c - t +  c_\psi \xt )}.
    \end{eqnarray}
\end{subequations}
We denote  $c_\psi = \cos \psi (t)$ and $s_\psi =  \sin \psi (t)$ and we have used the transform \eqref{tf} to express all spatial dependencies in terms of the $\wt{x}$-coordinate. For simplicity, we keep the notation $\xt$, $t$, $\zeta$, $\zeta_p$, and $h$ in this non-dimensional representation (although they actually represent the dimensional $k\xt$, $\omega t$, $k \zeta$, $k\zeta_p$ and $kh$). The non-dimensional problem that defines the strip deformation writes

\begin{eqnarray} \label{StripDeformation_Eq}
     \hat{l}_{{D}}^4 \frac{\partial^4\zeta_p}{\partial\xt^4} + \zeta_p & = & \zeta - \hat{\beta}\lz \quad  \text{with} \quad     \left.\frac{\partial^2\zeta_p}{\partial \xt^2}\right|_{\pm \lx/2} =   \left.\frac{\partial^3\zeta_p}{\partial\xt^3}\right|_{\pm \lx/2} = 0 .
    \label{StripDeformation_BC}
\end{eqnarray}
The solution to this problem provides access to the local submersion depth $h$ that appears in the the non-dimensional equations of motion for $x_c$ and $\psi$:

\begin{subequations}
    \begin{eqnarray}
        \ddot{x}_c& \approx & -  \frac{1 }{\hat{\beta} \lx \lz} \int_{-\lx/2}^{\lx/2}  \epsilon   e^{\zeta_p (\xt,t) } \, \cos{(x_c - t + c_\psi \xt )} \,  h(\xt,t)   \, d\xt \label{eq_mvt_xc}\\
        \ddot{\psi}  &\approx& \frac{12}{\hat{\beta} \lx^3 \lz}  \int_{-\lx/2}^{\lx/2} s_\psi \xt\, {\epsilon} e^{\zeta_p (\xt,t) } \, \cos{(x_c - t + c_\psi  \xt)}  \, h (\xt,t)  \,  d\xt . \label{eq_mvt_psi}
    \end{eqnarray}
\end{subequations}
Equations of motion \eqref{eq_mvt_xc} and \eqref{eq_mvt_psi}, together with the strip deformation equation \eqref{StripDeformation_Eq} and the boundary conditions of the strip deformation \eqref{StripDeformation_BC}, form the complete problem to solve.
We find an asymptotic solution to the previous system of integro-differential equations in the small wave limit $\epsilon \rightarrow 0$. Variables $x_c, \psi, \zeta_p $ are expanded in powers of $\epsilon$:

\begin{subequations}\label{EpsPowerSeries}
    \begin{eqnarray}
        x_c&=& \overline{x}_c + \epsilon x_c' + \epsilon^2 x_c'' \\
        \psi &=& \overline{\psi} + \epsilon \psi' + \epsilon^2 \psi'' \\
        \zeta_p & = & \overline{\zeta}_b + \epsilon \zeta_p ' + \epsilon^2 \zeta_p '' \\
        h & = & \hb+ \epsilon h' + \epsilon^2 h'' 
    \end{eqnarray}
\end{subequations}
We admit that variables can vary on multiple time-scales, meaning that 

\begin{equation}\label{EpsPowerSeries_time}
    \dot{\ } \rightarrow  \partial_t  + \epsilon \partial_\tau  +  \epsilon^2 \partial_T .
\end{equation}
We inject these expansions in the equations of motion \eqref{eq_mvt_xc} and \eqref{eq_mvt_psi}. 
In the integrands of  \eqref{eq_mvt_psi}, we also use the following Taylor expansions 

\begin{eqnarray}
    s_\psi &=& \sbpsi + \epsilon \psi'  \cbpsi + O (\epsilon^2) \nonumber \\
    \cos{(x_c - t + c_\psi  \xt)} &= & \cos{(\xbc - t + \cbpsi  \xt)} + \epsilon (- x_c'+ \psi'  \sbpsi \wt{x})  \sin{(\xbc - t + \cbpsi  \xt)}  +  O (\epsilon^2 )  \nonumber \\
    e^{\zeta_p } &=& e^{-\zetapb } ( 1 + \epsilon \zeta_p '  + O (\epsilon^2)  ). 
\end{eqnarray}
We denote $\sbpsi = \sin \psib$ and $\cbpsi = \cos \psib$. Here and further, we simplify  $e^{-\zetapb} \approx 1$ because by assumption $\hat{\beta} \lz \ll 1$.

\subsubsection{Perturbative analysis of the plate deformation}
We consider the case of a weakly deformable floater, such as those considered in section \ref{PrefOrientation_Expe}. In this limit, we can look for a perturbative solution of the thin plate equation \eqref{StripDeformation_Eq}. We divide this equation by $\hat{l}_{{D}}^{-4}$ :

\begin{eqnarray}
\frac{\partial^4\zeta_p}{\partial\xt^4} =   \hat{l}_{{D}}^{-4} \left ( \zeta - \zeta_p  -  \hat{\beta}\lz \right ) ,
\end{eqnarray}
and consider the limit $\hat{l}_{{D}}^{-4} \rightarrow 0$. The bending term dominates the equation and we propose a perturbative series solution

\begin{eqnarray}
\zeta_p = \zeta_p^{(0)} +  \hat{l}_{{D}}^{-4} \zeta_p^{(1)} + \hat{l}_{{D}}^{-8} \zeta_p^{(2)} + O( \hat{l}_{{D}}^{-12}).
\end{eqnarray}
The zero-th, first- and second-order problems are

\begin{eqnarray}
\frac{\partial^4\zeta_p^{(0)}}{\partial\xt^4} & = & 0,\label{Eq_zetap0}\\
\frac{\partial^4\zeta_p^{(1)}}{\partial\xt^4} +  \zeta_p^{(0)} & = &  \zeta - \hat{\beta} \lz,\label{Eq_zetap1}\\
\frac{\partial^4\zeta_p^{(2)}}{\partial\xt^4} +  \zeta_p^{(1)} & = &  0.\label{Eq_zetap2}
\end{eqnarray}
Given the boundary conditions in equations \eqref{StripDeformation_BC}, the first equation \eqref{Eq_zetap0} has solutions of the form $\zeta_p^{(0)} = A_1 \xt + A_0$, that need to be inserted in the second equation \eqref{Eq_zetap1} to find the constants $A_0$ and $A_1$. This yields

\begin{eqnarray}
\frac{\partial^4\zeta_p^{(1)}}{\partial\xt^4} = \zeta - A_1\xt - A_0 - \hat{\beta} \lz.
\end{eqnarray}
We denote $\mathcal{L} = \partial^4/\partial\xt^4$ the linear operator that appears in equations \eqref{Eq_zetap0} and \eqref{Eq_zetap1}. And we choose the $L^2$ inner product defined by 

\begin{eqnarray}
\langle \Psi |\mathcal{L} \phi\rangle = \int_{\lx/2}^{\lx/2}\Psi(\xt) (\mathcal{L}\phi)(\xt) d\xt.
\end{eqnarray}
With the boundary conditions \eqref{StripDeformation_BC}, we can show by successive integrations by parts that $\langle \Psi |\mathcal{L}\phi \rangle = \langle \mathcal{L}\Psi|\phi \rangle$, which means the operator $\mathcal{L}$ is self-adjoint. If we want the problem for $\zeta_p^{(1)}$ of \eqref{Eq_zetap1} to be solvable, we must demand that the right hand side is orthogonal to the kernel of $\mathcal{L}$, which is spanned by $\zeta_p^{(0)}$. This is called the solvability condition and it requires that

\begin{eqnarray}
\left\langle 1 \,\Big | \,\mathcal{L} \zeta_p^{(1)}\right\rangle = 0 \quad \text{and} \quad \left\langle \xt \,\Big | \,\mathcal{L} \zeta_p^{(1)}\right\rangle = 0,\label{SolvabilityCondition}
\end{eqnarray}
or more explicitly that 

\begin{subequations}
\begin{eqnarray}
 \int_{-\lx/2}^{\lx/2}\left (\zeta - A_1\xt - A_0 - \hat{\beta} \lz \right ) &=& 0 \\
  \int_{-\lx/2}^{\lx/2} \wt{x} \left (  \zeta - A_1\xt - A_0 - \hat{\beta} \lz \right ) &=& 0.
\end{eqnarray}
\end{subequations}
The first condition fixes the constant $A_0$, and the second condition fixes $A_1$. We Taylor expand the wave height $\zeta\simeq(1 - \cpsi^2\xt^2/2)\sin(x_c - t) + \cpsi \xt \cos(x_c - t) $ up to second order in $\lx$ and obtain

\begin{eqnarray}
\zeta_p^{(0)} = - \hat{\beta} \lz  + \epsilon \left(1 - \frac{ \overline{c}_\psi^2 \lx^2}{24}\right) \sin (x_c - t) + \epsilon  \overline{c}_\psi  \xt \cos(x_c - t)\label{eq_zetap0}
\end{eqnarray}
as leading order plate deformation, in the rigid plate limit. In this expression, the first term locates the floater's bottom face at equilibrium $z = - \hat{\beta} \lz = - \hb$.  The two other terms are contributions at order $\epsilon$. The first one is a uniform vertical displacement (bobbing) and the second one is a solid-body rotation (a rigid plate tries to align with the instantaneous surface). Notice that we find here the exact same expressions at order $\epsilon$ as in the rigid case without capillarity \citep{herreman2024}. This is a consequence of the hypothesis that the menisci are quasi-static, therefore only affecting the static immersion depth. 

To get the first order deformation of the plate $\zeta_p^{(1)}$, the solution of the order zero $\zeta_p^{(0)}$ is plugged in equation \eqref{Eq_zetap1}, which, integrated four times, yields

\begin{eqnarray}
\zeta_p^{(1)} & = & \epsilon \cpsi^2\left(\frac{l_x^2}{24} \frac{\xt^4}{4!} - \frac{\xt^6}{6!}\right)\sin(x_c - t) + B_3 \xt^3 + B_2 \xt^2 + B_1 \xt + B_0.
\end{eqnarray}
Here $B_0, B_1, B_2, B_3$ are four integration constants. The boundary conditions \eqref{StripDeformation_BC} set the values of $B_2$ and $B_3$, and $B_0$ and $B_1$ are fixed by the solvability conditions at second order:

\begin{eqnarray}
\left\langle 1 \,\Big | \,\mathcal{L} \zeta_p^{(2)}\right\rangle = 0 \quad \text{and} \quad \left\langle \xt \,\Big | \,\mathcal{L} \zeta_p^{(2)}\right\rangle = 0.\label{SolvabilityCondition_2}
\end{eqnarray}
We find 

\begin{subequations}
\begin{eqnarray}
   B_0 &=& \frac{29}{7!}\epsilon \cpsi^2 \frac{\lx^6}{2^6}\sin(x_c - t),\\
    B_1 &=& 0,\\
    B_2 &=& - \frac{1}{4!} \epsilon\cpsi^2 \frac{\lx^4}{2^5} \sin(x_c - t),\\
    B_3 & = & 0,
\end{eqnarray} 
\end{subequations}
and we write the plate deformation at order $\hat{l}_D^{-4}$ as

\begin{eqnarray}
\zeta_p^{(1)} & = & \epsilon \cpsi^2\left[\frac{29}{7!}\frac{l_x^6}{2^6} - \frac{l_x^4}{4!2^5}\xt^2 + \frac{l_x^2}{24}\frac{\xt^4}{4!} - \frac{\xt^6}{6!}\right] \sin(x_c - t).
\end{eqnarray}
For the remaining calculation, we can confuse $\zeta_p$ with $\zeta_p^{(0)} + \hat{l}_D^{-4} \zeta_p^{(1)} $ and  then we identify the zero and first order expression of our $\epsilon$-expansion of the plate surface:

\begin{eqnarray}
\overline{\zeta}_p = - \hb \quad \text{and} \quad \zeta_p' &= &\left(1 - \frac{\cpsi^2 \lx^2}{24}\right) \sin (x_c - t) + \cpsi \xt \cos(x_c - t)\nonumber\\
& + &  \cpsi^2\hat{l}_D^{-4}\left[\frac{29}{7!}\frac{l_x^6}{2^6} - \frac{l_x^4}{4!2^5}\xt^2 + \frac{l_x^2}{24}\frac{\xt^4}{4!} - \frac{\xt^6}{6!}\right] \sin(x_c - t).
\end{eqnarray}
The local immersion depth $h = \zeta - \zeta_p$ simplifies at first order in $\epsilon$ to

\begin{eqnarray}
    h(\xt,t)  =  \hb &+& \epsilon \cpsi^2 \left(\frac{\lx^2}{24} - \frac{\xt^2}{2}\right) \sin (x_c - t)\nonumber\\
 & - & \epsilon\cpsi^2\hat{l}_D^{-4}\left[\frac{29}{7!}\frac{l_x^6}{2^6} - \frac{l_x^4}{4!2^5}\xt^2 + \frac{l_x^2}{24}\frac{\xt^4}{4!} - \frac{\xt^6}{6!}\right] \sin(x_c - t).
\end{eqnarray}

\subsubsection{Mean yaw moment and preferential orientation}

The plate motion being known up to first order, we can now develop equations \eqref{eq_mvt_xc}-\eqref{eq_mvt_psi} for $x_c$ and $\psi$ in power series of $\epsilon$ using equations \eqref{EpsPowerSeries} and \eqref{EpsPowerSeries_time}. At order $\epsilon$, we have :

\begin{subequations}
    \begin{eqnarray}
        \partial^2_{tt} x_c'& = & -\frac{1 }{ \lx} \int \cos(x_c - t +  \overline{c}_\psi   \xt)   d\xt  ,\\
         \partial^2_{tt} \psi' &=& \frac{12}{  \lx^3} \int \spsi\xt\cos(x_c - t + \overline{c}_\psi  \xt)   d\xt  .
    \end{eqnarray}
\end{subequations}
We perform a Taylor expansion of the cosinus for a small floater $\lx \ll 1$, and find the leading order terms for the displacement $x_c'$ and yaw angle $\psi'$:

\begin{subequations}\label{xc'_psi'_rigid}
    \begin{eqnarray}
        x_c' & \approx & \cos(\xbc - t),\\
        \psi' & \approx & \overline{s}_\psi   \overline{c}_\psi   \sin(\xbc - t).
    \end{eqnarray}
\end{subequations}
We remark that equation \eqref{xc'_psi'_rigid}(a) is the first order motion of a fluid particle in the surface gravity waves. The first order motion is identical to the one without capillarity \citep{herreman2024}. 

To discuss preferential orientation, we must develop at order $\epsilon^2$ the equation for the yaw angle motion \eqref{eq_mvt_psi}:

\begin{eqnarray}\label{Eq_psi_epsilon2}
    \partial^2_{\tau\tau}\psib + \partial^2_{tt}\psi'' =  \frac{12}{\lx^3} \bigg\{&-&\int x'_c \left[\sbpsi \xt \sin(\xbc - t + \cbpsi\xt)\right]d\xt\nonumber\\
     &+& \int \psi' \left[\cbpsi\xt\cos(\xbc - t + \cbpsi\xt) + \sbpsi^2\xt^2 \sin(\xbc - t + \cbpsi\xt)\right]d\xt\nonumber\\
    &+&\int \zeta_p' \left [  \sbpsi \xt \cos(\xbc - t + \cbpsi\xt) \right ] d\xt\nonumber\\
    &+& \frac{1}{\overline{h}} \int h' \left [ \sbpsi\xt \cos(\xbc - t + \cbpsi\xt) \right ] d\xt\quad\bigg\}.
\end{eqnarray}
The first order motion $\zeta_p', x_c', \psi', h'$  is injected in equation \eqref{Eq_psi_epsilon2} and we also expand the trigonometric functions in powers of $\xt$, again because the floater is short ($l_x \ll 1$). Then, averaging over a wave period, we obtain the acceleration of the mean yaw angle as

\begin{eqnarray}\label{Mean_Yaw_Moment}
\partial^2_{\tau\tau} \psib\approx  - \sbpsi \cbpsi^3 \left(1 - \frac{\lx^2}{60\hb}\left(1 - \frac{1}{504}\left(\frac{l_x}{\hat{l}_D}\right)^4\right)\right).
\end{eqnarray}
It is this evolution equation that predicts the preferential orientation of the capillary floaters. For rigid floaters such that $l_x/\hat{l}_D\ll1$, this equation reduces to

\begin{eqnarray}\label{Mean_Yaw_Moment_Rigid}
\partial^2_{\tau\tau} \psib\approx  - \sbpsi \cbpsi^3 \left(1 - \frac{F}{60}\right), \quad\text{with}\quad F = \lx^2/\hb.
\end{eqnarray}
In the interval of interest $\overline{\psi} \in [0, 90^o]$, rigid floaters with $F= \lx^2/\hb < 60$ have a negative mean yaw moment and always rotate towards the longitudinal orientation $\overline{\psi} = 0^\circ$, whereas floaters with $F= \lx^2/\hb > 60$ have a positive mean yaw moment and rotate towards the transverse orientation $\overline{\psi} = 90^\circ$. Equation~\eqref{Mean_Yaw_Moment_Rigid} is identical to the expression derived in \cite{herreman2024} for rigid floaters without capillary effects, except that here the immersion depth $\hb$ is given by \eqref{hb} and includes capillary effects. This theoretical prediction agrees very well with the experimental data in figure~\ref{Fig4_Exp_F_vs_Lylc}.

For deformable capillary floaters, equation \eqref{Mean_Yaw_Moment} can be written as

\begin{eqnarray}\label{Mean_Yaw_Moment_ElastoCap}
\partial^2_{\tau\tau} \psib\approx  - \sbpsi \cbpsi^3 \left(1 - \frac{F}{F_c}\right), \quad \text{with} \quad F_c \approx 60 + \frac{5}{42}\left(\frac{l_x}{\hat{l}_D}\right)^4.
\end{eqnarray}
The critical value $F_c$ is no longer $60$ but depends on the ratio $l_x/\hat{l}_D$. This expression for $F_c$ is a good approximation of the expression derived in \cite{herreman2026elastic} for elastic floaters of arbitrary length without capillary effects, but with $\hb$ and $\hat{l}_D$ given in equations \eqref{hb} and \eqref{Adim} that include capillary effects. As anticipated in section \ref{rho_eff}, this is equivalent to using the effective density $\hat{\rho} = \rho(1 + 2\lc/\ly)$, hence confirming the critical value $F_c$ given in equation \eqref{Fc_elastocap_guess}. This theoretical prediction is plotted in figure~\ref{F_vs_LxlD_ManipsModel} as a dashed line and agrees well with the experimental data.

\subsection{Effects of tension and natural curvature}
\label{sec_natcurv}


In our model, we have neglected the tension term in equation \eqref{eq_plaque_gen} because the floaters in our experiments are only weakly deformable, with $L_D\gg\lc$. However for very flexible floaters, this term may significantly affect the equilibrium immersion of the plate. We show in the supplementary material how to solve the equation of the plate deformation \eqref{eq_plaque_gen} at equilibrium, including capillary and tension effects, for arbitrary flexural length. We find that tension terms only affect a small boundary layer near the tips of the floater. For the floaters considered in this article, the plate deformation at equilibrium differs by at most $7\%$ from the case without tension, confirming that this term can be neglected in equation \eqref{eq_plaque_gen}.

Another parameter neglected in the theory, which may have a significant influence in the experiments, is the natural curvature $\kappa$ of the floater.
Such curvature can be accounted for by changing the boundary conditions \eqref{StripDeformation_BC} to 

\begin{equation}\label{BC_NatCurv}
\left.\frac{\partial^2 \zeta_p}{\partial \tilde{x}^2}\right|_{\tilde{x} = \pm l_x/2} = -\kappa, \qquad \left.\frac{\partial^3\zeta_p}{\partial\tilde{x}^3}\right|_{\tilde{x} = \pm l_x/2} = 0.
\end{equation}
In the supplementary materials, we numerically solve the equilibrium, the first-order motion and the second-order mean yaw moment of the plate using this modified boundary condition.

\begin{figure}[tb]
    \centering
    \includegraphics[width = \linewidth]{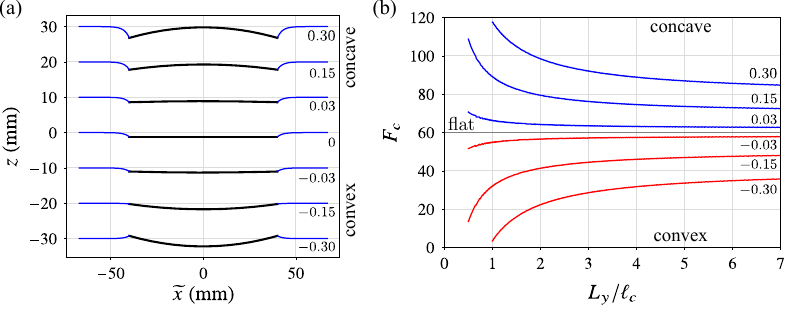}
    \caption{(a) Effect of the natural curvature (convex or concave) on the equilibrium shape of a capillary floater, for $L_x/L_D = 1.1$ and for curvatures $-0.3 \leq \kappa L_x \leq 0.3$. (b) Resulting critical value $F_c$ for the longitudinal-transverse transition, for concave and convex floaters of varying natural curvatures.}
    \label{AppendixANatCurvature}
\end{figure}

The effect of natural curvature is illustrated in figure~\ref{AppendixANatCurvature}(a), showing the equilibrium shapes for floaters such that $L_x = 80$~mm~$\simeq 1.1 L_D$, with varying natural curvatures $\kappa$. Positive values of $\kappa$ correspond to concave floaters, and negative values to convex floaters. Here $\kappa$ is chosen so that the height difference between the floater's centre and its tips is approximately a few millimetres, yielding values in the range $0.03\leq|\kappa| L_x\leq 0.3$.  We clearly see that the equilibrium immersion depth is affected by this natural curvature, suggesting that the preferential orientation should be affected as well. This is confirmed in figure~\ref{AppendixANatCurvature}(c), showing the transition line in the plan $(F,L_y/\lc)$ between the longitudinal and transverse orientations for convex and concave floaters, for various values of $\kappa$. Compared with flat floaters of the same size and density (for which $F_c=60$), concave floaters are more likely to align transversely, whereas convex floaters preferentially orient longitudinally. This shift in $F_c$ can be explained qualitatively as follows: first, a concave floater has a larger average submersion depth than its flat counterpart, resulting in a smaller threshold $F_c$, and hence a stronger preference for transverse orientation (and vice-versa for convex floaters); second, the tips of a concave floater are more immersed  than its flat counterpart, resulting in a stronger moment that pushes the floater toward the transverse orientation in wave troughs.


In experiments, if the brass plates are cut with scissors, this can induce a curvature on the order of $\kappa L_x \simeq0.15$. According to figure~\ref{AppendixANatCurvature}(c), the effects on the transition value $F_c$ can be as large as $50$\%. To ensure robust results on preferential orientation, it is therefore critical to minimize residual curvature effects, for instance by using a laser cutting technique.

\section{Conclusion}

In this paper we studied the preferential orientation of denser-than-water slender flexible floaters drifting in propagating gravity waves. We focused on capillary effects acting on floaters small compared to the wavelength. We theoretically derive a criterion to predict their preferential orientation, which generalizes the non-capillary rigid  and the non-capillary elastic  theories of \cite{herreman2024} and \cite{herreman2026elastic}, respectively. This generalised criterion shows excellent agreement with experiments carried out using brass rectangular plates of varying length, width, and thickness. These experiments also provide a validation of the non-capillary elastic theory of \cite{herreman2026elastic}, which could not be directly tested experimentally due to inherently non-negligible capillary effects in small-scale setups.

To account for capillary effects, we have shown that a generalised Archimedes principle including capillary forces, introduced by \cite{mansfield1997equilibrium} and \cite{keller1998surface} for equilibrium flotation, remains valid for our dynamic case of a small floater drifting in gravity waves, within the diffractionless Froude-Krylov approximation. When the waves are long compared to the capillary length and to the floater, the floater and the meniscus follow quasi-statically the wave motion.

We apply this quasi-static extension in our model of preferential orientation, together with hydroelastic effects using the Kirchoff-Love equation for thin plates. This elasto-capillary model leads to a remarkably simple result: the preferential orientation of a capillary floater is governed by the same non-dimensional number $F$ as in \cite{herreman2024} and \cite{herreman2026elastic}, by including an effective fluid density $\hat \rho$ that incorporates capillary effects. 

Our approach can be extended to floaters of arbitrary shape, as long as their local radius of curvature is everywhere large compared to the capillary length, or if the singularities of the meniscus (induced by sharp corners) can be neglected in the equilibrium flotation -- which is precisely our assumption in the case of slender plates. In this case, the capillary forces reduce to a contour integral along the floater edge of the local fluid acceleration, with the difficulty of modelling the varying meniscus angle along the contour. Combining this with the pressure forces acting on the floater's wetted surface, the force and moment reduce to a volume integral over the total displaced fluid, with an effective fluid density in the form 

$$
\hat \rho \simeq \rho \left(1 + \ell_c \frac{\mbox{perimeter}}{\mbox{wetted surface}} \right)
$$
(for a slender rectangle with $L_x \gg L_y$, the ratio surface/perimeter is $L_y/2$). This approach provides a practical correction for floater-wave interaction problems where capillary effects cannot be neglected \citep{ho2023capillary,harris2025propulsion}.

Although the present study focuses on propagating gravity waves, the specific nature of the waves is not critical to the model. In particular, the assumption of purely gravity  waves is not essential, as the framework remains applicable to capillary-gravity waves, provided the characteristic size of the object remains sufficiently small. This level of generality enables extension of the model to other prescribed wave field profiles under similar size constraints. Consequently, the model also offers a suitable basis for investigating the drift of a small particle in a standing wave, as examined by \cite{falkovich2005floater}.

\vspace{4mm}

{\small

\noindent {\bf Acknowledgements.} We thank  A. Aubertin, L. Auffray, J. Amarni, S. Mergui and R. Pidoux for experimental help. This work was supported by the project “TransWaves” (Project No. ANR-24-CE51-3840-01) of the French National Research Agency.

\vspace{4mm}

\noindent {\bf Declaration of interests.} The authors report no conflict of interest.

}

\bibliographystyle{jfm}   
\bibliography{jfm}

\end{document}